\documentclass[prl,twocolumn,
nofootinbib,tightenlines,superscriptaddress]{revtex4}

\usepackage{times}
\usepackage{natbib}
\usepackage{amssymb,amsbsy,amsmath,amsfonts}
\usepackage{mathrsfs}
\usepackage{graphicx}% Include figure files
\usepackage{epsf,epsfig,float,latexsym,amsthm,fancyhdr,rotating}
\usepackage{graphics,psfrag,longtable}
\usepackage{slashed}
\usepackage{esint}
\usepackage{nicefrac}

\DeclareMathAlphabet{\mathantt}{OML}{antt}{l}{it}
\DeclareMathAlphabet{\mathpzc}{OT1}{pzc}{m}{n}

\def\beq{\begin{equation}}
\def\eeq{\end{equation}}
\def\bea{\begin{eqnarray}}
\def\eea{\end{eqnarray}}
\def\beqa{\begin{equation}\begin{array}{l}}
\def\eeqa{\end{array}\end{equation}}
\def\barr{\left(\begin{array}{c}}
\def\earr{\end{array}\right)}
\def\bmat{\left(\begin{array}{cc}}
\def\emat{\end{array}\right)}
\graphicspath{{./images/}}

\begin{document}
%\preprint{MITP}
\title {Low energy doubly-virtual Compton scattering from di-lepton electroproduction on a nucleon}

\author{Vladyslav Pauk}

\affiliation{Institut f\"ur Kernphysik, Cluster of Excellence PRISMA$^+$,  Johannes Gutenberg-Universit\"at, D-55099 Mainz, Germany}

\author{Carl E. Carlson}

\affiliation{College of William and Mary, Physics Department, Williamsburg, Virginia 23187, USA}

\author{Marc Vanderhaeghen}

\affiliation{Institut f\"ur Kernphysik, Cluster of Excellence PRISMA$^+$,  Johannes Gutenberg-Universit\"at, D-55099 Mainz, Germany}

\begin{abstract}
We propose a new way to experimentally determine the subleading low-energy structure constant of doubly-virtual Compton scattering on a proton. Such empirical determination will reduce the theoretical model error in estimates of the hadronic correction to the muonic hydrogen Lamb shift. We demonstrate that the di-lepton forward-backward asymmetry in the $e^- p \to e^- p \, e^- e^+$ process, which can be accessed at electron scattering facilities, yields a large sensitivity to this so far unknown low-energy constant.

\end{abstract}
\date{\today}
\maketitle
%\thispagestyle{empty}
%\tableofcontents

Extractions of the proton charge radius from muonic hydrogen ($\mu$H) Lamb shift measurements over the past decade~\cite{Pohl:2010zza,Antognini:1900ns} have reported a highly precise proton-radius value, with more than an order of magnitude improvement in the precision. These results disagreed by around 5.6 standard deviations, with the values obtained from measurements of energy level shifts in electronic hydrogen~\cite{Mohr:2015ccw} or from electron-proton elastic scattering experiments~\cite{Bernauer:2010wm}. This so-called ''proton radius puzzle" has spurred a lot of activity, 
see e.g.~\cite{Pohl:2013yb,Carlson:2015jba} for reviews. 
A new round of experiments using electronic hydrogen spectroscopy~\cite{Beyer:2017,Bezginov:2019mdi}, as well as a 
new electron scattering experiment~\cite{Xiong:2019umf} are favoring the lower value of the radius consistent with the $\mu$H spectroscopy results, although a recent electronic hydrogen experiment~\cite{Fleurbaey:2018fih} has also 
reported a large value of the proton charge radius. To fully clarify this situation, further experiments both with electron beams~\cite{Denig:2016tpq}, 
muon beams~\cite{Gilman:2017hdr,Denisov:2018unj}, or by a direct comparison of cross sections for 
$\gamma\ p\rightarrow e^+e^-p$ versus $\gamma\ p\rightarrow \mu^+\mu^-p$~\cite{Pauk:2015oaa}, are presently planned or underway. 
With the next generation of high-precision experiments both in scattering and spectroscopy, 
the focus is now shifting to improve the precision on this fundamental nucleon structure quantity, as spelled out recently in~\cite{Hammer:2019uab}.  
Indeed, to extract the proton charge radius from $\mu$H Lamb shift measurements, which is at present the most precise method, 
the proton form factors, structure functions and polarizabilities are all required as input in a quantitative understanding of the hadronic 
correction~\cite{Carlson:2011zd, Birse:2012eb, Antognini:2013jkc}. At present the theoretical uncertainty due to this correction, 
which is evaluated in a dispersive framework, is of the same size as the 2P - 2S $\mu$H Lamb shift experimental uncertainty,  
and is the main limitation when converting a value of the Lamb shift  
to a value for the proton radius. 
%The uncertainty is amplified further when considering muonic deuterium ($\mu$D) or muonic $^3$He$^+$. There, the experimental uncertainty is already a factor 5 or 6 smaller as compared to the uncertainty due to the hadronic corrections. 

The main part of this hadronic uncertainty results from the subtraction function entering the forward doubly-virtual Compton scattering process. It corresponds to the situation where the photons in the Compton process have zero energy and finite virtuality. At second order in the photon virtuality, this function is constrained by the magnetic polarizability, which is determined experimentally~\cite{Tanabashi:2018oca}.     
To fourth order in the photon virtuality, one low-energy constant in this subtraction function is at present empirically unconstrained~\cite{Lensky:2017bwi}, and one relies on chiral effective field theory calculations~\cite{Birse:2012eb, Alarcon:2013cba} or phenomenological estimates.  
We demonstrate in this work that this low-energy constant  
can be accessed experimentally through the forward-backward asymmetry in the $e^- p \to e^- p \, e^- e^+$ process. 
This observable is directly sensitive to the interference between the QED and doubly-virtual Compton amplitudes. 
A pioneering measurement of this forward-backward asymmetry in the $\gamma p \to e^- e^+ p$ process has been performed at DESY quite some time ago~\cite{Alvensleben:1973mi} as a test of the Kramers-Kronig relation at high energies. In the present work, we demonstrate that the corresponding experiments with a spacelike initial virtual photon, which can be realized at electron scattering facilities, yield a large sensitivity to this so far unknown low-energy constant. 

The helicity averaged forward doubly-virtual Compton scattering process (VVCS), $\gamma^\ast(q) + N(p) \to \gamma^\ast(q) + N(p)$ is described by two invariant amplitudes, denoted by $T_1$ and $T_2$, which are functions of two kinematic invariants: $Q^2 = -q^2$ and $\nu = q \cdot p / M$, with $M$ the nucleon mass. 
Its covariant tensor structure in the four-vector indices of initial ($\mu$) and final ($\nu$) photons can be written, following notations from~\cite{Pasquini:2018wbl}, as:
\bea
\label{vvcs}
\alpha_\mathrm{em}   M^{\mu \nu} (\mathrm{VVCS}) 
\equiv \hat g^{\mu \nu} T_1(\nu, Q^2) 
- \frac{\hat p^{\mu}  \hat p^{\nu}}{M^2}  T_2 (\nu, Q^2) , 
\eea
with $\hat g^{\mu \nu} \equiv g^{\mu\nu}- q^{\mu}q^{\nu} / q^2$, $\hat p^\mu \equiv p^{\mu}- p\cdot q / q^2 \,q^{\mu}$, and 
where $\alpha_\mathrm{em} = e^2 / 4 \pi \simeq 1/137$.  
The optical theorem relates the imaginary parts of $T_1$ and $T_2$ as:
\bea
\label{optical}
{\rm{Im}}\ T_1(\nu, Q^2) = \frac{e^2}{4M} F_1  \, , \quad
{\rm{Im}}\ T_2(\nu, Q^2)  = \frac{e^2}{4 \nu} F_2  \, ,  
\eea
where $F_1, F_2$ are the conventionally defined structure functions parametrizing 
inclusive electron-nucleon scattering, and depend on $Q^2$ and $x \equiv Q^2 / 2 M \nu$.  
The two-photon exchange correction to the $\mu$H Lamb shift can be expressed as a weighted double integral over $Q^2$ and $\nu$ 
of the forward amplitudes $T_1$ and $T_2$~\cite{Carlson:2011zd}. Using the empirical input of $F_1$ and $F_2$, the  $\nu$ dependence 
of $T_2$ has been fully reconstructed in ~\cite{Carlson:2011zd} using an unsubtracted dispersion relation, whereas the dispersion relation for $T_1$ requires one subtraction, which can be chosen at $\nu = 0$ as $T_1(0, Q^2)$. 
The subtraction function is usually split in a Born part (corresponding with the nucleon intermediate state), and a remainder, so-called non-Born part, which we denote in the following by $\bar T_1(0, Q^2)$. Although the Born part can be expressed in terms of elastic form factors (see  \cite{Pasquini:2018wbl} for the corresponding expression), the non-Born part cannot be fixed empirically so far. 
In general, one can however write down a low $Q^2$ expansion of $\bar T_1(0, Q^2)$ as:
\bea
\bar T_1(0, Q^2) = \beta_{M1} Q^2 + \frac{1}{2} T_1^{''}(0) Q^4 + \mathcal{O}(Q^6),
\label{T1bar0}
\eea
where the term proportional to $Q^2$ is empirically determined by the magnetic dipole polarizability $\beta_{M1}$~\cite{Tanabashi:2018oca}.  
Theoretical estimates for the subtraction term were given at order $Q^4$ in heavy-baryon chiral perturbation theory (HBChPT)~\cite{Birse:2012eb}, and 
covariant baryon chiral perturbation theory (BChPT), both at leading order (LO) due to $\pi N$ loops,
and at next-to-leading order (NLO), including both $\Delta(1232)$-exchange and $\pi \Delta$ loops~\cite{Alarcon:2013cba,Lensky:2017bwi}. 
Furthermore, estimates of the subtraction function  
were also extracted from superconvergence sum rule (SR) relations, relying on a dispersion relation for the difference of the structure function $F_1$  and a Regge fit of its high-energy behavior~\cite{Tomalak:2015hva}. The different values obtained for $\bar{T}_1^{\prime\prime}(0)$  are compared in Table~\ref{tab:subtraction} (second column). Even for these theoretically well motivated approaches, the spread among the different estimates is quite large. The resulting uncertainty due to this subtraction term constitutes at present the main uncertainty in the theoretical Lamb shift estimate. 
We next discuss how to avoid such model dependence, and propose an empirical way to determine $\bar{T}_1^{\prime\prime}(0)$. 
\begin{table}[h]
\begin{tabular}{cc|c|c}
\hline\hline
Source & Ref. & $\frac{1}{2}\bar{T}_1^{\prime\prime}(0)$ & $\alpha_\mathrm{em}  b_{3,0}$ 
\\
\hline
HBChPT & \cite{Birse:2012eb} & $[-1.01, -0.35]$ &  \\
\hline
$\pi N$ loops &
& $-0.06$  & $0.001$ \\
%\hline
$\pi \Delta$ loops &
& $-0.10$ & $-0.005$ \\
%\hline
$\Delta$ exchange &
& $-1.98$ & $0.11$ \\
Total BChPT & \cite{Lensky:2017bwi}
& $-2.14 \pm 0.98$ & $0.11 \pm 0.05$  \\
\hline
superconvergence SR & \cite{Tomalak:2015hva} 
& $-0.47$ & $3.96$  \\
\hline\hline
\end{tabular}
\caption{Values of the $Q^4$ term of the subtraction function $\bar T_1(0,Q^2)$ (second column) and of the dVCS low-energy constant $b_{3,0}$ (third column), both in units $10^{-4}$ fm$^5$, in different theoretical approaches~\cite{Lensky:2017bwi}. The indicated range for the HBChPT result corresponds with the range given by Eq.~(15) in Ref.~\cite{Birse:2012eb}.
\label{tab:subtraction}
}
\end{table}

To empirically access the $Q^4$ term and potentially also higher order terms in $\bar T_1(0, Q^2)$, 
we consider the full off-forward doubly-virtual Compton process, 
  $\gamma^\ast(q) + N(p) \to \gamma^\ast(q^\prime) + N(p^\prime)$
, where both photons are virtual. 
In the following, we study the case where the initial photon is spacelike ($q^2 < 0$), and the final photon is timelike ($q^{\prime 2} > 0$), which can be accessed experimentally. 
In general, the full off-forward doubly-virtual Compton scattering (dVCS) amplitude off a proton 
is described by 18 tensor structures  in the initial ($\mu$) and final ($\nu$) photon four-vector indices~\cite{Tarrach:1975tu}. 
In this work,  we will only need the  
helicity-averaged amplitude,   
which is described by 5 independent  tensors, and can be expressed as~\cite{Drechsel:1997xv}:
\bea
M^{\mu \nu} &=& 
\sum_{i =1,2,3,4,19} B_i(\nu, q^2, q^{\prime 2}, q \cdot q^\prime) \, T_i^{\mu \nu} , 
\label{eq:dvcsunpol}
\eea
where $T_i^{\mu \nu}$ are the spin-independent and gauge invariant tensors, symmetric under exchange of the two virtual photons, 
see Eq.~(8) in \cite{Lensky:2017bwi} for the corresponding expressions.
Furthermore, in~(\ref{eq:dvcsunpol}), the invariant amplitudes $B_i$ are functions of four Lorentz invariants, 
with $\nu \equiv q \cdot P / M$, where $P \equiv (p + p^\prime) / 2$.    
As the forward VVCS process of Eq.~(\ref{vvcs}) is a special case of Eq.~(\ref{eq:dvcsunpol}), one can express the subtraction function 
as~\cite{Lensky:2017bwi}:
\bea
\bar T_1(0, Q^2) = \alpha_\mathrm{em} Q^2 \left( \bar B_1 + Q^2  \bar B_3   \right),
\eea
where both non-Born amplitudes $\bar B_{1}, \bar B_3$ are understood in the forward limit ($q = q^\prime$), 
i.e. $\bar B_{i}(0, q^2, q^2, q^2)$ for $i = 1,3$.

The Born contribution was worked out in Ref.~\cite{Tarrach:1975tu}. For the helicity averaged amplitude, it only contributes to 
the amplitudes $B_1$ and $B_2$, see Eq.~(8) in \cite{Lensky:2017bwi}.
The non-Born part of the dVCS amplitudes, denoted as $\bar B_i$, can be expanded 
for small values of $q^2, q^{\prime \, 2}, q \cdot q^\prime$ and $\nu$, 
with coefficients given by polarizabilities. As the 
$\nu$ dependence can be fully reconstructed up to at most one subtraction, through a dispersion relation, we only need to discuss the 
amplitudes entering the subtraction function. To determine $\bar T_1(0,Q^2)$ up to the $Q^4$ term,   
we use the low-energy expansion in $k \in \{q, q^\prime\}$~\cite{Lensky:2017bwi}:
\bea
\bar{B}_1(0, q^2, q^{\prime 2}, q \cdot q^\prime) &=& \frac{1}{\alpha_\mathrm{em}} \left\{ \beta_{M1} - \frac{1}{6} \beta_{M2} q \cdot q^\prime 
\right. \nonumber \\
&&\left. \hspace{-1.5cm}- \left(\beta^\prime_{M1}(0) + \frac{\beta_{M1}}{8 M^2}  \right) ( q^2 + q^{\prime \, 2}) \right\} +  {\cal  O}(k^4), \nonumber \\ 
\bar{B}_3(0, q^2, q^{\prime 2}, q \cdot q^\prime)  &=& b_{3,0} + {\cal  O}(k^2), 
\label{lexdvcs}
\eea
where $\beta_{M2}$ is the magnetic quadrupole polarizability determined from real Compton scattering~\cite{Holstein:1999uu}, 
and $\beta^\prime_{M1}(0)$ is the slope at $Q^2 = 0$ of the generalized magnetic dipole polarizability which is accessed through 
virtual Compton scattering, see Ref.~\cite{Fonvieille:2019eyf} for a recent review. The low-energy constant $b_{3, 0}$ is not determined empirically so far because the tensor structure $T_3^{\mu \nu}$ decouples when either the initial or final photon is real.
Using Eq.~(\ref{lexdvcs}), we can express the $Q^4$ term in Eq.~(\ref{T1bar0}) as:
\bea
\frac{1}{2} T_1^{''}(0) = \frac{1}{6} \beta_{M2} + 2 \beta^\prime_{M1}(0) + \frac{\beta_{M1}}{4 M^2} +  \alpha_\mathrm{em} b_{3,0}. 
\eea
We compare in Table~\ref{tab:subtraction} (third column)  several theoretically motivated estimates for $b_{3,0}$. 
We see that the BChPT including $\Delta$-pole corresponds with a very small value of $b_{3,0}$ in comparison with the value resulting from the superconvergence SR estimate for $T^{\prime \prime}_1(0)$.  

To empirically determine $\bar{T}_1^{\prime\prime}(0)$ and $b_{3,0}$ 
we consider the process of electroproduction of a di-lepton (electron-positron) pair on the nucleon,
\bea 
e^- (k) + N (p) \rightarrow e^- (k^\prime) + N (p^\prime) + e^- (l_-) + e^+(l_+), 
\label{eq:eNeNll}
\eea
where the four-momenta of the corresponding particles are shown in parentheses.
 We define the eight-fold phase space of the reaction (\ref{eq:eNeNll}) in terms of five invariants: 
 \bea
 s &=& (k + p)^2, \quad \quad Q^2 = -(k - k^\prime)^2, \quad \quad W^2 = (q + p)^2, \nonumber \\
 t &=&(p^\prime - p)^2,  \hspace{0.52cm} q^{\prime \, 2} = (l_- + l_+)^2,
 \eea
 and three angles $\Phi$, $\theta_l$, and $\phi_l$. 
The invariant $s$ is obtained from the electron beam energy $E_e$ as $s = M^2 + 2 M E_e$, 
 $\Phi$ is the angle of the intial electron plane relative to the production plane, spanned by the vectors $q \equiv k - k^\prime$ and $q^\prime \equiv l_- + l_+$ in the c.m. frame ($\vec q + \vec p = 0$), and $\theta_l$,  $\phi_l$ are the angles of the produced negative lepton in the di-lepton rest frame (with polar angle defined relative to the c.m. direction of $q^\prime$). 
 The differential cross section of the reaction (\ref{eq:eNeNll}) reads
 \begin{align}
&\frac{d \sigma}{dQ^2 dW^2 d\Phi dt dq^{\prime 2} (d\Omega_l)_{e^-e^+}}= \frac{1}{(4\pi)^7 } \frac{1}{2 (s-M^2)^2} \nonumber\\
& \hspace{2.cm}\times \frac{(1 - 4 m_e^2 / q^{\prime 2})^{1/2} }{\lambda(W^2,M^2,-Q^2)^{1/2}} \overline{\sum_i}\sum_f |{\cal M}|^2,
\end{align}
where $m_e$ is the lepton mass, 
$\lambda$ is the K\"all\'en triangle function, and $\cal M$ stands for the amplitude of the reaction (\ref{eq:eNeNll}).
\begin{figure}[h]
  \includegraphics[width=8cm]{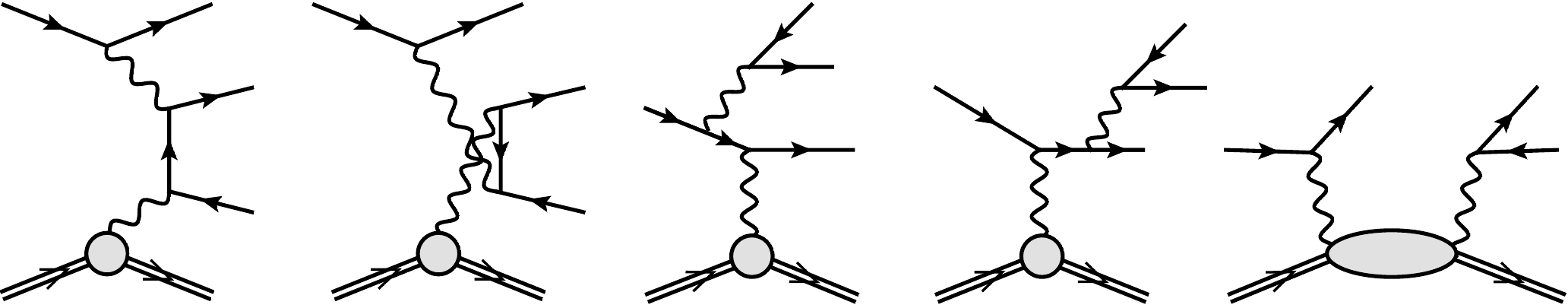}
  \caption{Feynman diagrams for the spacelike BH process (1 and 2), the timelike BH process (3 and 4), 
  and the dVCS process (diagram 5). The single (double) lines correspond with leptons (nucleons).}
  \label{fig:bh}
\end{figure}
At ${\cal{O}}(\alpha_{QED}^2)$, the reaction (\ref{eq:eNeNll}) is described by the processes shown in Fig.~\ref{fig:bh}. 
The first four diagrams correspond with the spacelike (SL) and timelike (TL) Bethe-Heitler (BH) processes, whereas the last is the dVCS process. 
The BH-SL and BH-TL diagrams are fully determined by the nucleon's electromagnetic FFs, represented by the blobs in diagrams 1 - 4 in 
Fig.~\ref{fig:bh}. We adopt the FF parameterization of~\cite{Bernauer:2010wm} in the following, which we can analytically continue to the 
small timelike virtualities considered here.  
The invariant amplitude for the dVCS process (${\cal M}_{C}$), where initial (final) photons have  spacelike (timelike) virtualities, is given by 
\begin{equation}
{\cal M}_{C} =   \frac{i  e^4}{q^{\prime  2} Q^2}  
\bar N(p^\prime) M^{\mu \nu} N(p) \bar u(k^\prime) \gamma_\mu u(k)
 \bar u(l_-) \gamma_\nu v(l_+), \nonumber 
\label{eq:MdVCS}
\end{equation}
where $M^{\mu\nu} $ is the dVCS tensor of Eq.~(\ref{eq:dvcsunpol}). 
To access the real part of the dVCS amplitude, with the aim to empirically extract the low-energy constant $b_{3,0}$ 
we consider in the following the forward-backward asymmetry $A_{FB}$, which is defined in the di-lepton rest frame as:
\bea
A_{FB} \equiv \frac{\overline{\sum_i} \sum_f   \left\{ | {\cal M} |^2_{\theta_l, \phi_l} -  |{\cal M} |^2_{\pi - \theta_l, \phi_l + \pi}  \right\} }{\overline{\sum_i} \sum_f   \left\{ | {\cal M} |^2_{\theta_l, \phi_l} +  |{\cal M} |^2_{\pi - \theta_l, \phi_l + \pi} \right\}}.
\label{eq:fb}
\eea 
The only non-zero contribution to this observable comes from the interference between processes with an even (BH-SL) and odd (BH-TL and dVCS)  number of photon couplings to the di-lepton pair, due to charge conjugation. 
Explicitly, the numerator in Eq.~(\ref{eq:fb}) is proportional to $\Re \left[ {\cal M}_{BH-SL} \left( {\cal M}_{BH-TL} + {\cal M}_{C} \right)^\ast \right]$. 
\begin{figure}[h]
 \includegraphics[width=4.25cm]{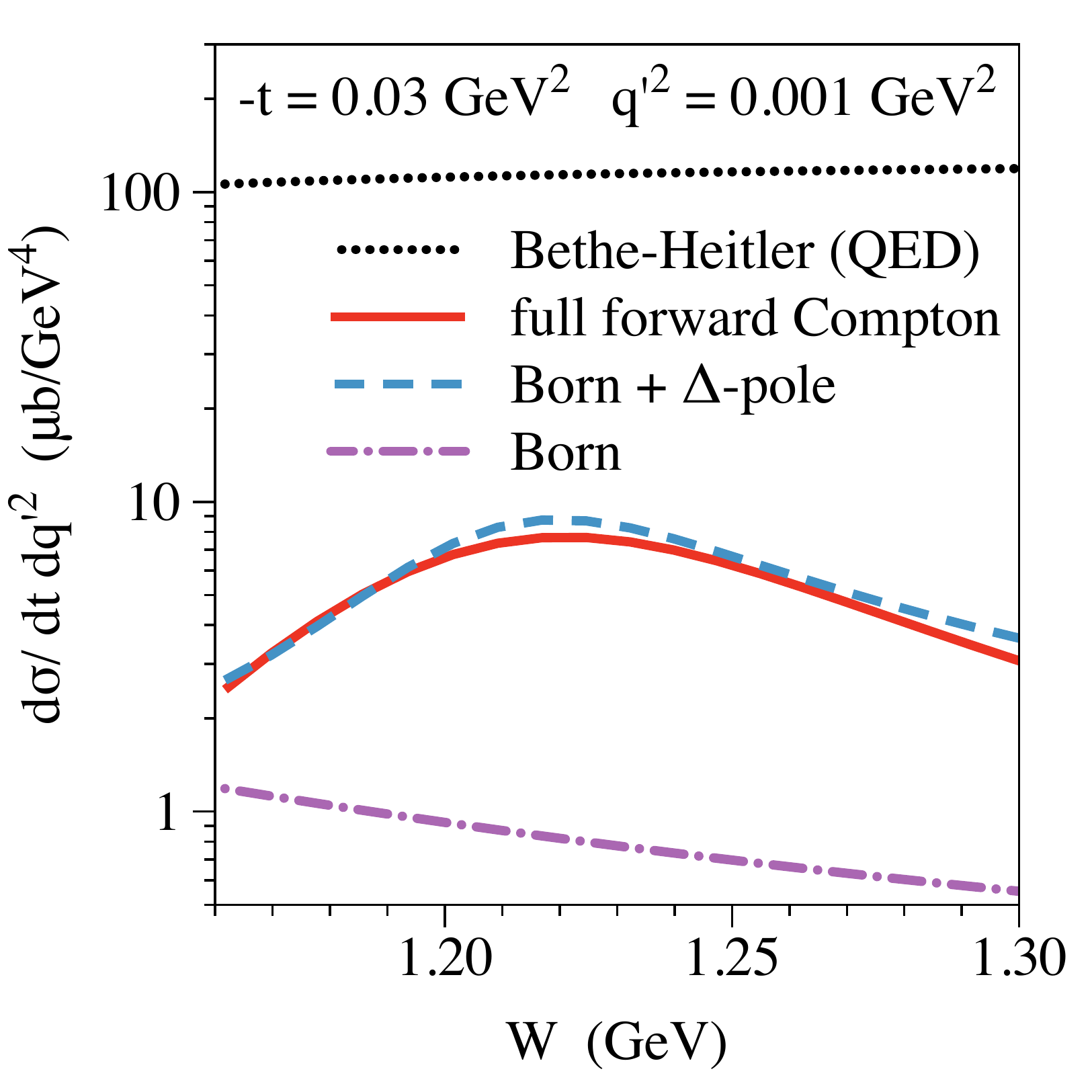}
   \includegraphics[width=4.25cm]{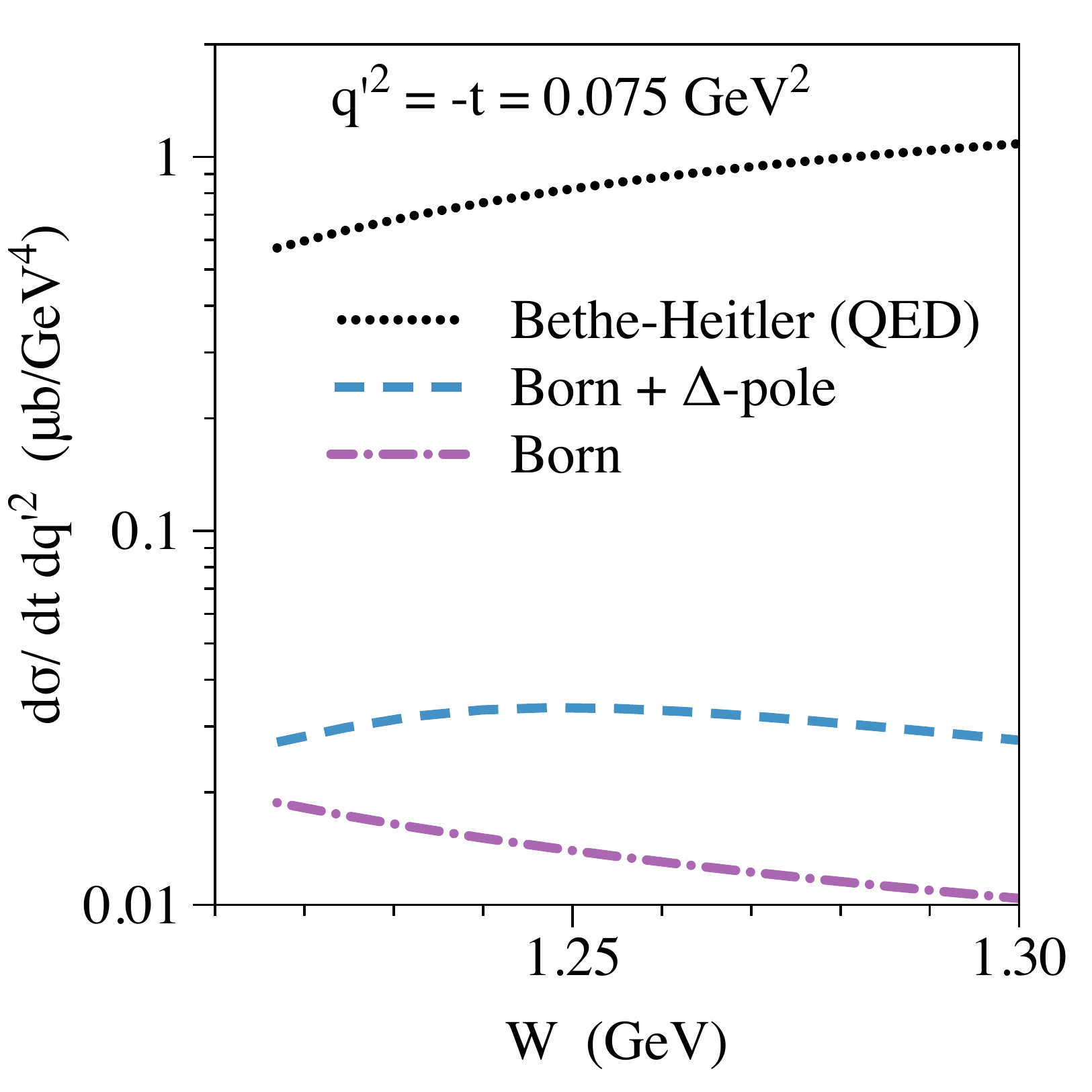}
  \caption{W-dependence of the $\gamma p \to  e^- e^+ p$ cross section integrated over the di-lepton angles 
  for two kinematic settings. The BH cross section is compared with 
  different models for the Compton cross section. A full calculation based on the empirical forward structure functions~\cite{Gryniuk:2015eza} is also shown for the near-forward quasi-real Compton kinematics (left panel, solid red curve).}
  \label{fig:cross_tcs}
\end{figure}

\begin{figure}[h]
  \includegraphics[width=6.5cm]{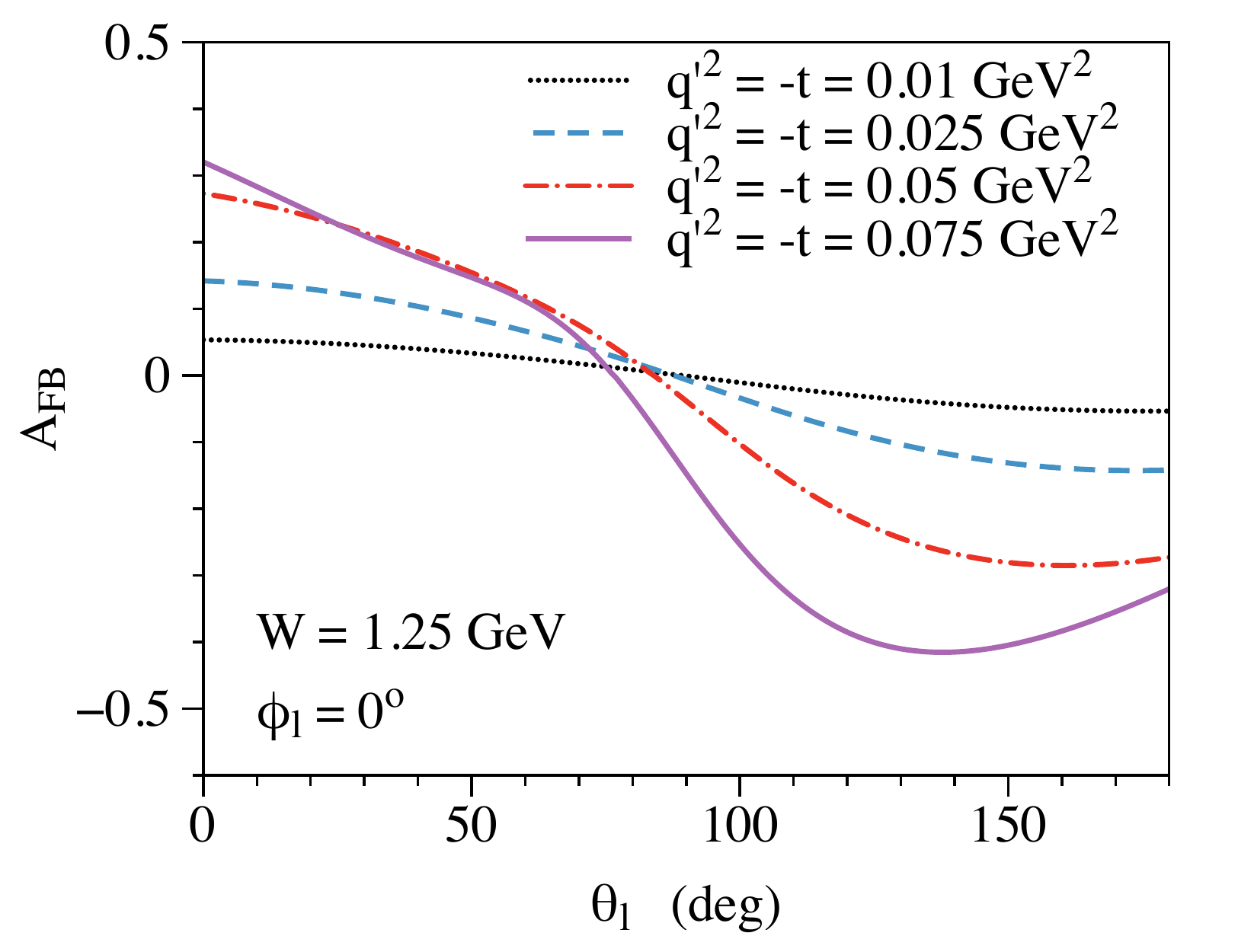}
  \caption{Lepton angular dependence (in di-lepton c.m. frame, for $\phi_l = 0^o$) of the $\gamma p \to e^- e^+ p$  
  asymmetry $A_{FB}$ for $W = 1.25$~GeV, and for different values of $q^{\prime \, 2}$ and $-t$. }
  \label{fig:AFB_tcs}
\end{figure}

We start our discussion by considering the case of an initial real photon through the $\gamma p \to e^- e^+ p$ reaction.  
In Fig.~\ref{fig:cross_tcs} we show the dependence of the $\gamma p \to e^- e^+ p$ cross section on the c.m. energy $W$ for two settings. One of  kinematics is approaching the forward real Compton process (left panel), with $q^{\prime \, 2}$ and $t$ values near the ones considered in the experiment of~\cite{Alvensleben:1973mi}.   
To provide a model of the inelastic effects in the dVCS process we consider an effective description of the non-Born part of the dVCS amplitude in the $\Delta(1232)$ region by the $\Delta$-pole amplitude. For the electromagnetic $N \to \Delta$ transition we use the empirical parameterization, see~\cite{Pascalutsa:2006up}. To  estimate the accuracy of the description, we also implemented for the near-forward real Compton situation, with $-t$ small and $q^{\prime \, 2}$ very close to zero, a full dispersive calculation based on empirical structure functions~\cite{Gryniuk:2015eza}. The latter calculation was found to be consistent with the so far only data point for $A_{FB}$~\cite{Alvensleben:1973mi}. We see from Fig.~\ref{fig:cross_tcs} that around c.m. energy $W = 1.25$~GeV the Compton part of the cross section integrated over the di-lepton solid angle is reproduced by the  
Born + $\Delta$-pole description within an accuracy of 5\% or better. Furthermore, 
for larger values of $q^{\prime \, 2} = -t$, the ratio between BH and Compton cross sections decreases.
Therefore, one expects an increase of the asymmetry $A_{FB}$ 
with increasing values of $q^{\prime \, 2}$ and $-t$, as is demonstrated in the lepton angular dependence in 
Fig.~\ref{fig:AFB_tcs} for $W = 1.25$~GeV. One sees that for $q^{\prime \, 2} = -t = 0.075$~GeV$^2$, $A_{FB}$ reaches values 
between -40\% and +30\%.

\begin{figure}[h]
 \includegraphics[width=6.5cm]{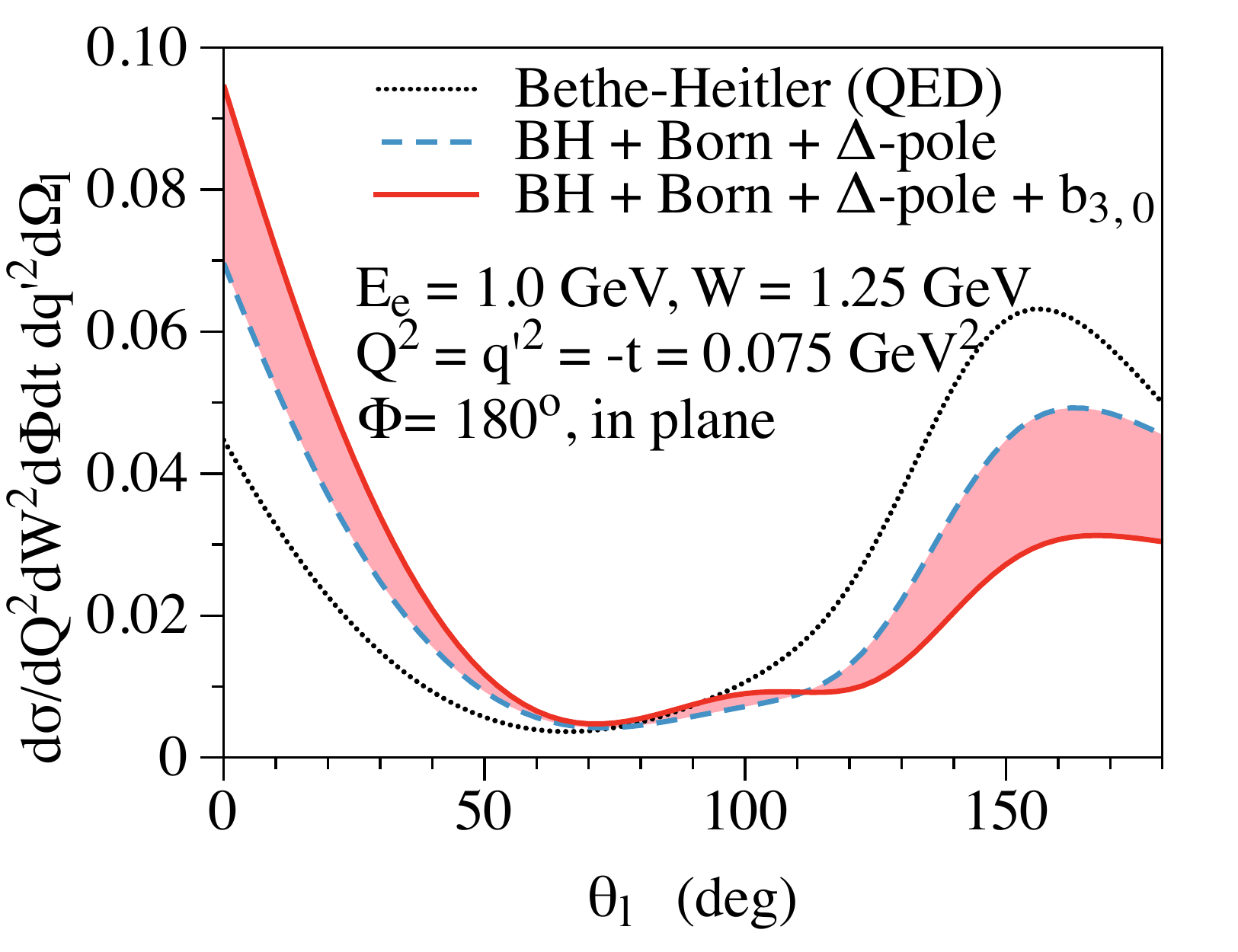}
   \includegraphics[width=6.5cm]{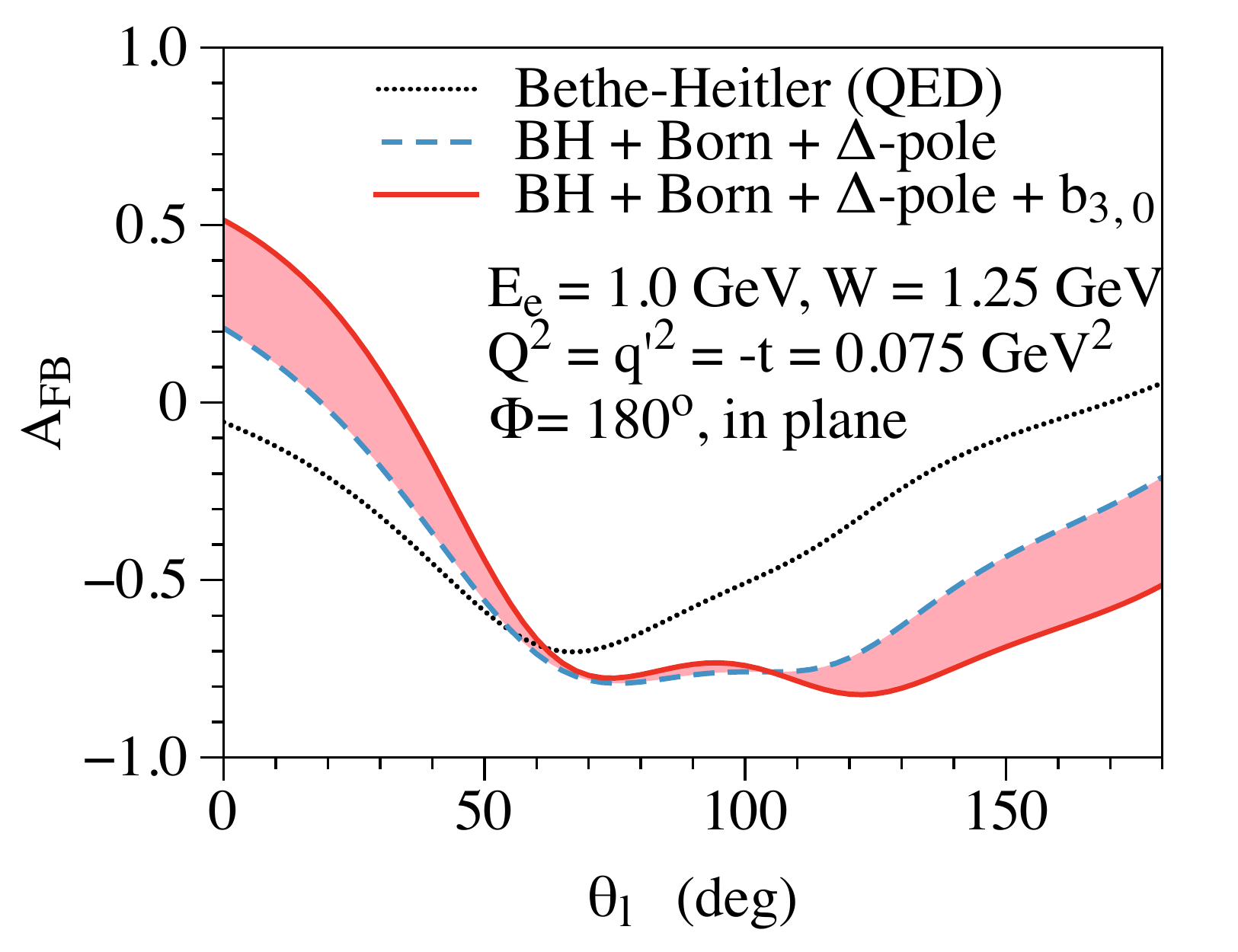}
  \caption{Lepton angular dependence (in di-lepton c.m. frame) of the $e^- p \to e^- p \, e^- e^+$  
  differential cross section (upper panel, in units nb/GeV$^8$sr$^2$) and asymmetry $A_{FB}$ (lower panel). 
%  for $E_e = 1$~GeV, $W = 1.25$~GeV, and $Q^2 = q^{\prime \, 2} = -t = 0.075$~GeV$^2$. 
The red band shows the sensitivity on the 
  low-energy constant $b_{3, 0}$, corresponding with the spread in Table~\ref{tab:subtraction}.}
  \label{fig:AFB_0p75}
\end{figure}

Having assessed the sensitivity of $A_{FB}$ to the dVCS amplitude for real photons, we next extend it to the initial virtual photon case through the $e^- p \to e^- p \, e^- e^+$  reaction. As $A_{FB}$ depends on the real part of the dVCS amplitude, it holds promise to study the sensitivity on the low-energy constant $b_{3,0}$.   
In Fig.~\ref{fig:AFB_0p75}, we show the cross section and $A_{FB}$ for the $e^- p \to e^- p \, e^- e^+$ process at $W = 1.25$~GeV, where the $\Delta$-pole was found to yield a very good description of the total Compton result, and in kinematics where $Q^2 = q^{\prime \, 2} = -t = 0.075$~GeV$^2$. 
One notices that in the forward and backward angular ranges the Bethe-Heitler process yields only a small asymmetry.  
In these ranges, the Compton process yields a large change in the asymmetry, up to 50\%. The red band shows the sensitivity on $b_{3, 0}$, corresponding with the spread in Table~\ref{tab:subtraction}. For forward and backward angles, where the sensitivity is the largest, the spread in the values for the subtraction constant $T_1^{\prime \prime}(0)$ 
in Table~\ref{tab:subtraction}, corresponds with a change in $A_{FB}$ from 20\% to 50\%.

In Summary, we have explored a direct empirical determination of the subtraction function in the doubly-virtual Compton process on a nucleon, which corresponds at present with the leading uncertainty in the hadronic correction to muonic atom spectroscopy. To fourth order in the photon virtuality, one low-energy quantity in the subtraction function is so far empirically unconstrained. We have demonstrated that it 
can be accessed experimentally through the forward-backward asymmetry of the $e^- p \to e^- p \, e^- e^+$ process. 
This observable is directly sensitive to the interference between the QED and dVCS amplitudes. 
Different theoretical estimates for this low-energy constant induce a change between 20\% and 50\% in the corresponding asymmetry in the $\Delta(1232)$ region.  This observable can be accessed by precision experiments at the electron facilities MAMI, MESA, and JLab. 
\\
\\
This work was supported by the Deutsche Forschungsgemein-schaft (DFG, German Research Foundation),
in part through the Collaborative Research Center [The Low-Energy Frontier of the Standard
Model, Projektnummer 204404729 - SFB 1044], and through the Cluster of Excellence
[Precision Physics, Fundamental Interactions, and Structure~of~Matter] (PRISMA+ EXC
2118/1) within the German Excellence Strategy (Project ID 39083149). 
CEC thanks the National Science Foundation (USA) for support under grants PHY-1516509 and  PHY-1812326, and the Johannes Gutenberg-University, and the Nordic Institute for Theoretical Physics (NORDITA) for hospitality while this work was underway.

\end{document}